\documentclass[5p,twocolumn,12pt]{elsarticle}

\usepackage{amssymb}

\usepackage{textcomp,graphicx,times,wrapfig,xspace,url,amssymb,amsmath,wasysym,stmaryrd}

\journal{Astroparticle Physics}

\begin{document}
\begin{frontmatter}


\title{The drive system of the\\Major Atmospheric Gamma-ray Imaging Cherenkov Telescope}

\author[tb]{T.~Bretz\corref{cor1}}
\author[tb]{D.~Dorner}
\author[rw]{R.~M.~Wagner}
\author[rw]{P.~Sawallisch}
\address[tb]{Universit\"{a}t W\"{u}rzburg, Am Hubland, 97074 W\"{u}rzburg, Germany}
\address[rw]{Max-Planck-Institut f\"ur Physik, F\"ohringer Ring 6, 80805 M\"{u}nchen, Germany}
\cortext[cor1]{Corresponding author: tbretz@astro.uni-wuerzburg.de}

\newcommand{\mylesssim}{{\apprle}}
\newcommand{\mygtrsim} {{\apprge}}
\newcommand{\degree}{{\textdegree{}}}

\begin{abstract}
The MAGIC telescope is an imaging atmospheric Cherenkov telescope,
designed to observe very high energy gamma-rays while achieving a low
energy threshold. One of the key science goals is fast follow-up of the
enigmatic and short lived gamma-ray bursts. The drive system for the
telescope has to meet two basic demands: (1)~During normal
observations, the 72-ton telescope has to be positioned accurately, and
has to track a given sky position with high precision at a typical
rotational speed in the order of one revolution per day. (2)~For
successfully observing GRB prompt emission and afterglows, it has to be
powerful enough to position to an arbitrary point on the sky within a
few ten seconds and commence normal tracking immediately thereafter. To
meet these requirements, the implementation and realization of the
drive system relies strongly on standard industry components to ensure
robustness and reliability. In this paper, we describe the mechanical
setup, the drive control and the calibration of the pointing, as well
as present measurements of the accuracy of the system. We show that the
drive system is mechanically able to operate the motors with an
accuracy even better than the feedback values from the axes. In the
context of future projects, envisaging telescope arrays comprising
about 100 individual instruments, the robustness and scalability of the
concept is emphasized.
\end{abstract}


\end{frontmatter}


\section{Introduction}

The MAGIC telescope on the Canary Island of La~Palma, located 2200\,m
above sea level at 28\textdegree{}45$^\prime$\,N and 17\textdegree{}54$^\prime$\,W, is
an imaging atmospheric Cherenkov telescope designed to achieve a low
energy threshold, fast positioning, and high tracking
accuracy~\cite{Lorenz:2004, Cortina:2005}. The MAGIC design, and the
currently ongoing construction of a second telescope
(MAGIC\,II;~\cite{Goebel:2007}), pave the way for ground-based
detection of gamma-ray sources at cosmological distances down to less
than 25\,GeV~\cite{Sci}. After the discovery of the distant blazars 1ES\,1218+304
at a redshift of $z$\,=\,0.182~\citep{2006ApJ...642L.119A} and
1ES\,1011+496 at $z$\,=\,0.212~\citep{2007ApJ...667L..21A}, the most
recent breakthrough has been the discovery of the first quasar at very
high energies, the flat-spectrum radio source 3C\,279 at a redshift of
$z$\,=\,0.536~\cite{2008Sci...320.1752M}. These observational results
were somewhat surprising, since the extragalactic background radiation
in the mid-infrared to near-infrared wavelength range was believed to
be strong enough to inhibit propagation of gamma-rays across
cosmological distances~\citep{2001MNRAS.320..504S, 2007arXiv0707.2915K, Hauser:2001}. 
The
apparent low level of pair attenuation of gamma-rays greatly improves
the prospects of searching for very high energy gamma-rays from
gamma-ray bursts (GRBs), cf.~\citep{Kneiske:2004}. Their remarkable similarities with blazar
flares, albeit at much shorter timescales, presumably arise from the
scaling behavior of relativistic jets, the common physical cause of
these phenomena. Since most GRBs reside at large redshifts, their
detection at very high energies relies on the low level of
absorption~\citep{1996ApJ...467..532M}. Moreover, the cosmological
absorption decreases with photon energy, favoring MAGIC to discover
GRBs due to its low energy threshold.

Due to the short life times of GRBs and the limited field of view of
imaging atmospheric Cherenkov telescopes, the drive system of the MAGIC
telescope has to meet two basic demands: during normal observations,
the 72-ton telescope has to be positioned accurately, and has to
track a given sky position, i.e., counteract the apparent rotation of
the celestial sphere, with high precision at a typical rotational speed
in the order of one revolution per day. For catching the GRB prompt
emission and afterglows, it has to be powerful enough to position the
telescope to an arbitrary point on the sky within a very short time
and commence normal tracking immediately
thereafter. To keep the system simple, i.e., robust, both requirements
should be achieved without an indexing gear. The telescope's total
weight of 72~tons is comparatively low, reflecting the
use of low-weight materials whenever possible.
For example, the mount consists of a space frame of carbon-fiber
reinforced plastic tubes, and the mirrors are made of polished
aluminum.

In this paper, we describe the basic properties of the MAGIC drive
system. In section~\ref{sec2}, the hardware components and mechanical
setup of the drive system are outlined. The control loops and
performance goals are described in section~\ref{sec3}, while the
implementation of the positioning and tracking algorithms and the
calibration of the drive system are explained in section~\ref{sec4}.
The system can be scaled to meet the demands of other telescope designs
as shown in section~\ref{sec5}. Finally, in section~\ref{outlook} and
section~\ref{conclusions} we draw conclusions from our experience of
operating the MAGIC telescope with this drive system for four years.

\section{General design considerations}\label{design}

The drive system of the MAGIC telescope is quite similar to that of
large, alt-azimuth-mounted optical telescopes. Nevertheless there are
quite a few aspects that influenced the design of the MAGIC drive
system in comparison to optical telescopes and small-diameter Imaging
Atmospheric Cherenkov telescopes (IACT).

Although IACTs have optical components, the tracking and stability
requirements for IACTs are much less demanding than for optical
telescopes. Like optical telescopes, IACTs track celestial
objects, but observe quite different phenomena: Optical telescopes
observe visible light, which originates at infinity and is parallel.
Consequently, the best-possible optical resolution is required and in
turn, equal tracking precision due to comparably long integration
times, i.e., seconds to hours. In contrast, IACTs record the Cherenkov
light produced by an electromagnetic air-shower in the atmosphere,
induced by a primary gamma-ray, i.e., from a close by
(5\,km\,-\,20\,km) and extended event with a diffuse transverse
extension and a typical extension of a few hundred meters. Due to the
stochastic nature of the shower development, the detected light will
have an inherent limitation in explanatory power, improving normally
with the energy, i.e., shower-particle multiplicity. As
the Cherenkov light is emitted under a small angle off the particle
tracks, these photons do not even point directly to the source like in
optical astronomy. Nevertheless, the shower points towards the
direction of the incoming gamma-ray and thus towards its source on the
sky. For this reason its origin can be reconstructed analyzing its
image. Modern IACTs achieve an energy-dependent pointing resolution for
individual showers of 6$^\prime$\,-\,0.6$^\prime$. These are the
predictions from Monte Carlo simulations assuming, amongst other
things, ideal tracking. This sets the limits achievable in practical
cases. Consequently, the required tracking precision must be at least
of the same order or even better. Although the short integration times,
on the order of a few nanoseconds, would allow for an offline
correction, this should be avoided since it may give rise to an
additional systematic error.


To meet one of the main physics goals, the observation of prompt and
afterglow emission of GRBs, positioning of the telescope to their
assumed sky position is required in a time as short as possible.
Alerts, provided by satellites, arrive at the MAGIC site typically
within 10\,s after the outburst~\citep{2007ApJ...667..358A}. 
Since the life times of GRBs show a bimodal distribution~\cite{Paciesas:1999}
with a peak between 10\,s and 100\,s. To achieve
a positioning time to any position on the sky within a reasonable
time inside this window, i.e. less than a minute,
a very light-weight but sturdy telescope and a fast-acting and
powerful drive system is required.
\begin{figure*}[htbp]
\begin{center}
 \includegraphics*[width=0.91\textwidth,angle=0,clip]{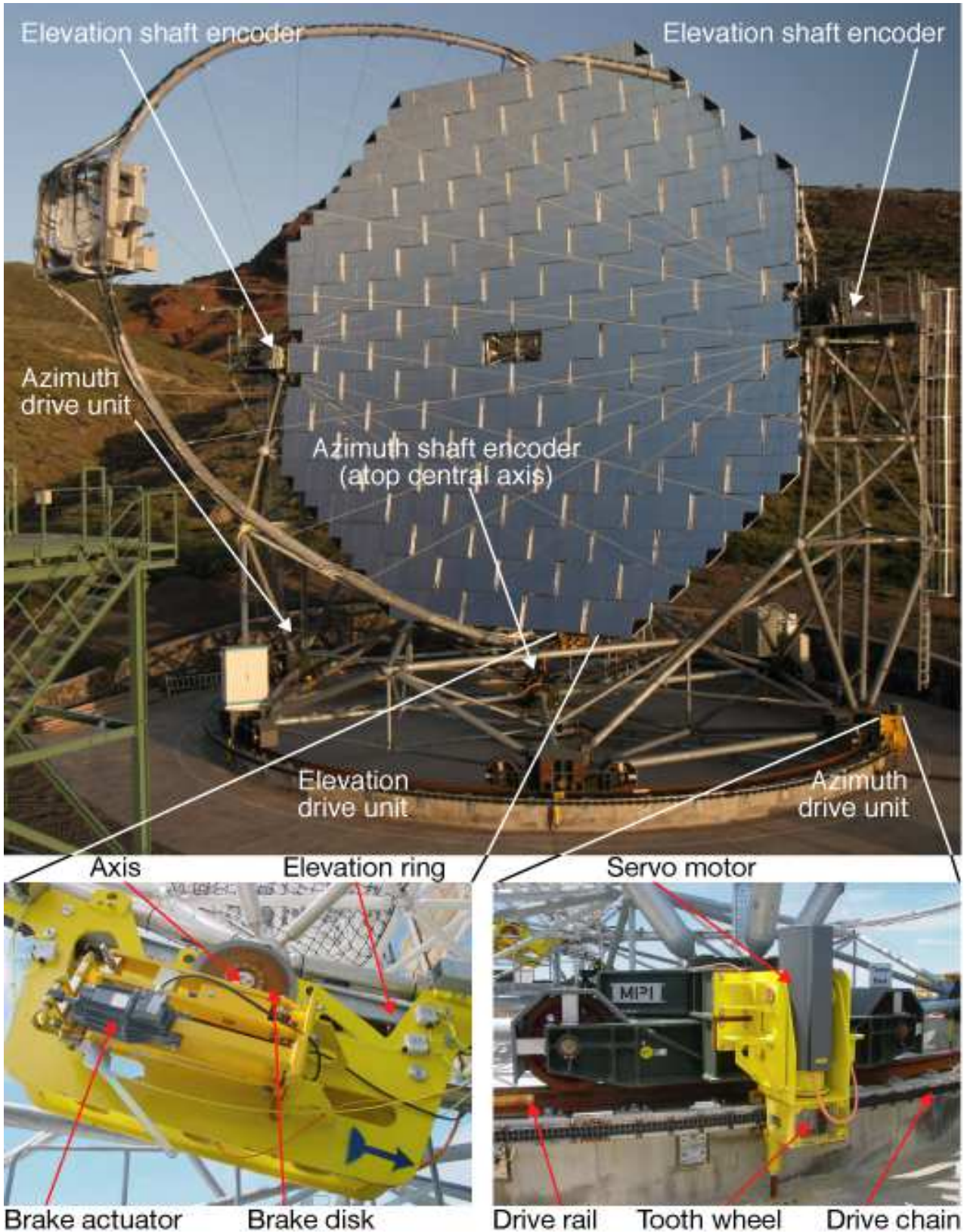}
\caption{The top picture shows the MAGIC\,I telescope with the major
components of the drive system. The elevation drive unit, from its back
side, is enlarged in the bottom left picture. Visible is the actuator
of the safety holding brake and its corresponding brake disk mounted on
the motor-driven axis. The motor is attached on the opposite side. The
picture on the bottom right shows one of the azimuth bogeys with its
two railway rails. The motor is housed in the grey box on the
yellow drive unit. It drives the tooth double-toothed wheel gearing
into the chain through a gear and a clutch.}
\label{figure1}
\end{center}
\end{figure*}

\section{Mechanical setup and hardware components}\label{sec2}

The implementation of the drive system relies strongly on standard
industry components to ensure robustness, reliability and proper
technical support. Its major drive components, described hereafter, are
shown on the pictures in fig.~\ref{figure2}. 

The azimuth drive ring of 20\,m diameter is made from a normal railway
rail, which was delivered in pre-bent sections and welded on site. Its
head is only about 74\,mm broad and has a bent profile. The fixing onto
the concrete foundation uses standard rail-fixing elements, and allows
for movements caused by temperature changes. The maximum allowable
deviation from the horizontal plane as well as deviation from flatness
is $\pm 2$\,mm, and from the ideal circle it is $\Delta$r\,=\,8\,mm. 
The rail support was leveled with a theodolite every 60\,cm
with an overall tolerance of $\pm 1.5$\,mm every 60\,cm. In between the
deviation is negligible. Each of the six bogeys holds two standard
crane wheels of 60\,cm diameter with a rather broad wheel tread of
110\,mm. This allows for deviations in the 11.5\,m-distance to the
central axis due to extreme temperature changes, which can even be
asymmetric in case of different exposure to sunlight on either side.
For the central bearing of the azimuth axis, a high-quality ball
bearing was installed fairly assuming that the axis is vertically
stable. For the elevation axis, due to lower weight, a less expensive
sliding bearing with a teflon layer was used. These sliding bearings
have a slightly spherical surface to allow for small misalignments
during installation and some bending of the elevation axis stubs under
load. 

The drive mechanism is based on duplex roller chains and sprocket
wheels in a rack-and-pinion mounting. The chains have a breaking
strength of 19~tons and a chain-link spacing of 2.5\,cm. The initial
play between the chain links and the sprocket-wheel teeth is about
3\,mm\,-\,5\,mm, according to the data sheet, corresponding to much
less than an arcsecond on the telescope axes. The azimuth drive chain
is fixed on a dedicated ring on the concrete foundation, but has quite
some radial distance variation of up to 5\,mm. The elevation drive
chain is mounted on a slightly oval ring below the mirror dish, because
the ring forms an integral part of the camera support mast structure.



Commercial synchronous motors (type designation Bosch
Rexroth\footnote{\url{http://www.boschrexroth.de}\\Bosch Rexroth AG,
97816 Lohr am Main, Germany} MHD\,112C-058) are used together with
low-play planetary gears (type designation
alpha\footnote{\url{http://www.wittenstein-alpha.de}\\Wittenstein alpha
GmbH, 97999 Igersheim, Germany} GTS\,210-M02-020\,B09, ratio 20) linked
to the sprocket wheels. These motors intrinsically allow for a
positional accuracy better than one arcsecond of the motor axis. Having
a nominal power of 11\,kW, they can be overpowered by up to a factor
five for a few seconds. It should be mentioned that due to the
installation height of more than 2200\,m a.s.l., due to lower air
pressure and consequently less efficient cooling, the nominal values
given must be reduced by about 20\%. Deceleration is done operating the
motors as generator which is as powerful as acceleration. The motors
contain 70\,Nm holding brakes which are not meant to be used as driving
brake. The azimuth motors are mounted on small lever arms. In order to
follow the small irregularities of the azimuthal drive chain, the units
are forced to follow the drive chain, horizontally and  vertically, by
guide rolls. The elevation-drive motor is mounted on a nearly 1\,m long
lever arm to be able to compensate the oval shape of the chain and the
fact that the center of the circle defined by the drive chain is
shifted 356\,mm away from the axis towards the camera. The elevation
drive is also equipped with an additional brake, operated only as
holding brake, for safety reasons in case of extremely strong wind
pressure. No further brake are installed on the telescope.

The design of the drive system control, c.f.~\citet{Bretz:2003drive},
is based on digitally controlled industrial drive units, one for each
motor. The two motors driving the azimuth axis are coupled to have a
more homogeneous load transmission from the motors to the structure
compared to a single (more powerful) motor. The modular design allows
to increase the number of coupled devices dynamically if necessary,
c.f.~\citet{Bretz:2005drive}.

At the latitude of La Palma, the azimuth track of stars can exceed
180\textdegree{} in one night. To allow for continuous observation of a
given source at night without reaching one of the end positions in
azimuth. the allowed range for movements in azimuth spans from
$\varphi$\,=\,-90\textdegree{} to $\varphi$\,=\,+318\textdegree, where
$\varphi$\,=\,0\textdegree{} corresponds to geographical North, and
$\varphi$\,=\,90\textdegree{} to geographical East. To keep slewing
distances as short as possible (particularly in case of GRB alerts),
the range for elevational movements spans from
$\theta$\,=\,+100\textdegree{} to $\theta$\,=\,-70\textdegree{} where
the change of sign implies a movement {\em across the zenith}. This
so-called {\it reverse mode} is currently not in use, as it might result
in hysteresis effects of the active mirror control system, still under
investigation, due to shifting of weight at zenith. The accessible
range in both directions and on both axes is limited by software to the
mechanically accessible range. For additional safety, hardware end
switches are installed directly connected to the drive controller
units, initiating a fast, controlled deceleration of the system when
activated. To achieve an azimuthal movement range exceeding
360\textdegree{}, one of the two azimuth end-switches needs to be
deactivated at any time. Therefore, an additional {\em direction
switch} is located at $\varphi$\,=\,164\textdegree{}, short-circuiting
the end switch currently out of range.

\section{Setup of the motion control system}\label{sec3}

\begin{figure*}[htb]
\begin{center}
 \includegraphics*[width=0.48\textwidth,angle=0,clip]{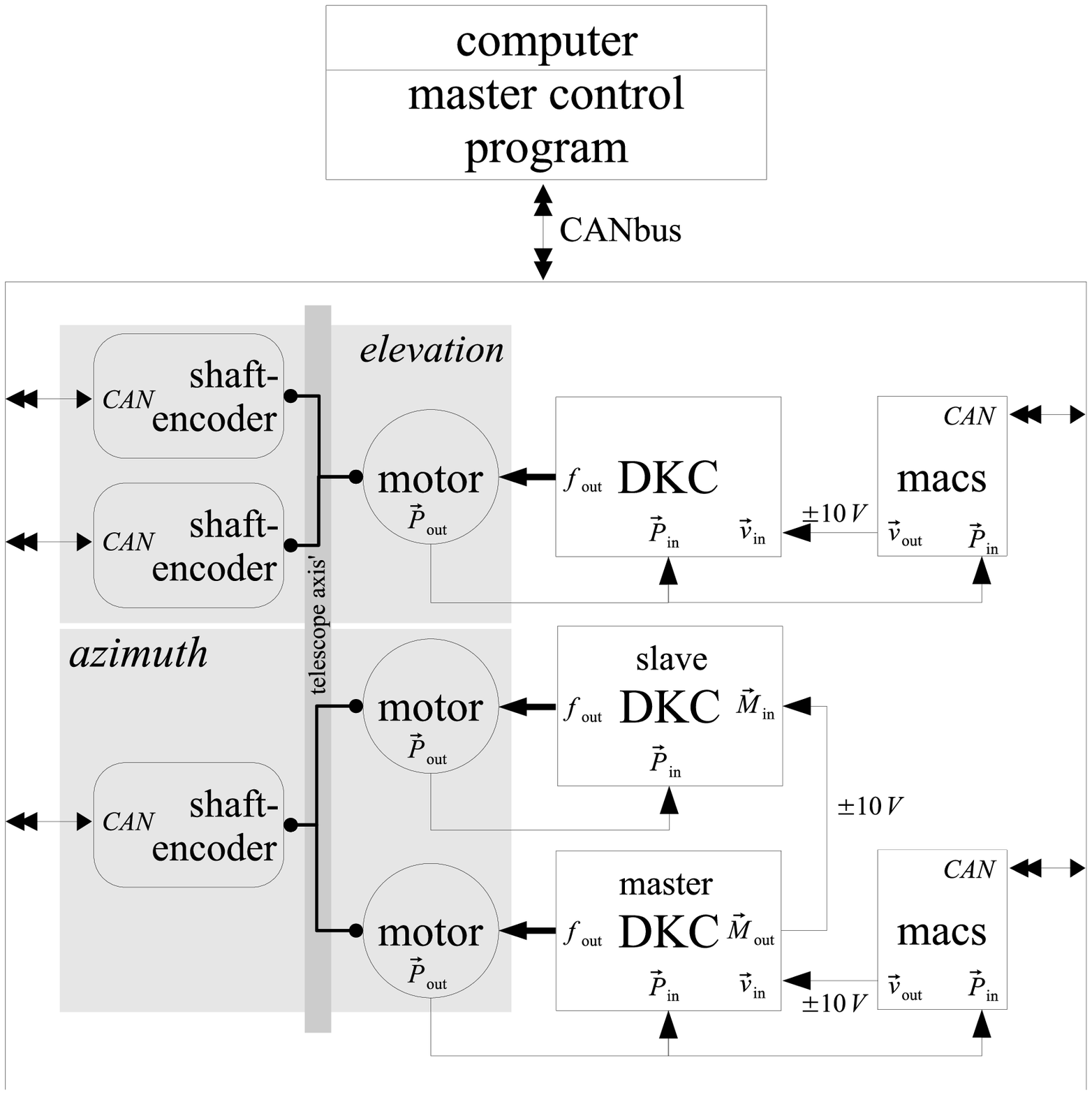}
 \hfill
 \includegraphics*[width=0.48\textwidth,angle=0,clip]{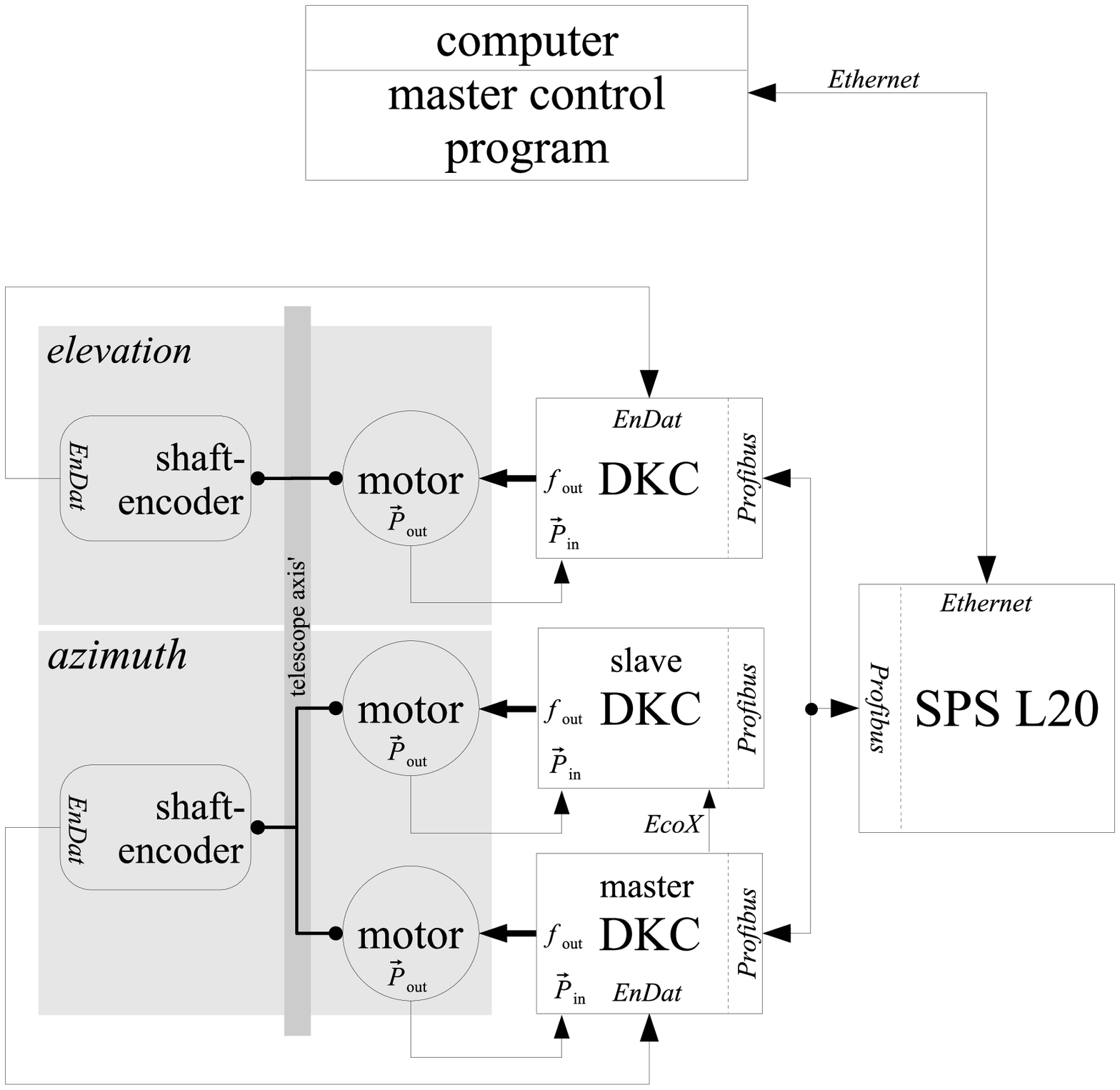}
\caption{Schematics of the MAGIC\,I ({\em left}) and MAGIC\,II ({\em
right}) drive system. The sketches shows the motors, the motor-encoder
feedback as well as the shaft-encoder feedback, and the motion-control
units, which are themselves controlled by a superior control, receiving
commands from the control PC, which closes the position-control loop.
The system is described in more details in section~\ref{sec3}.}
\label{figure2}
\end{center}
\end{figure*}

The motion control system similarly uses standard industry components.
The drive is controlled by the feedback of encoders measuring the
angular positions of the motors and the telescope axes. The encoders on
the motor axes provide information to micro controllers dedicated for
motion control, initiating and monitoring every movement. Professional
built-in servo loops take over the suppression of oscillations. The
correct pointing position of the system is ensured by a computer
program evaluating the feedback from the telescope axes and initiating
the motion executed by the micro controllers. Additionally, the
motor-axis encoders are also evaluated to increase accuracy. The
details of this system, as shown in figure~\ref{figure2}, are discussed
below.

\subsection{Position feedback system}

\begin{figure}[htb]
\begin{center}
 \includegraphics*[width=0.48\textwidth,angle=0,clip]{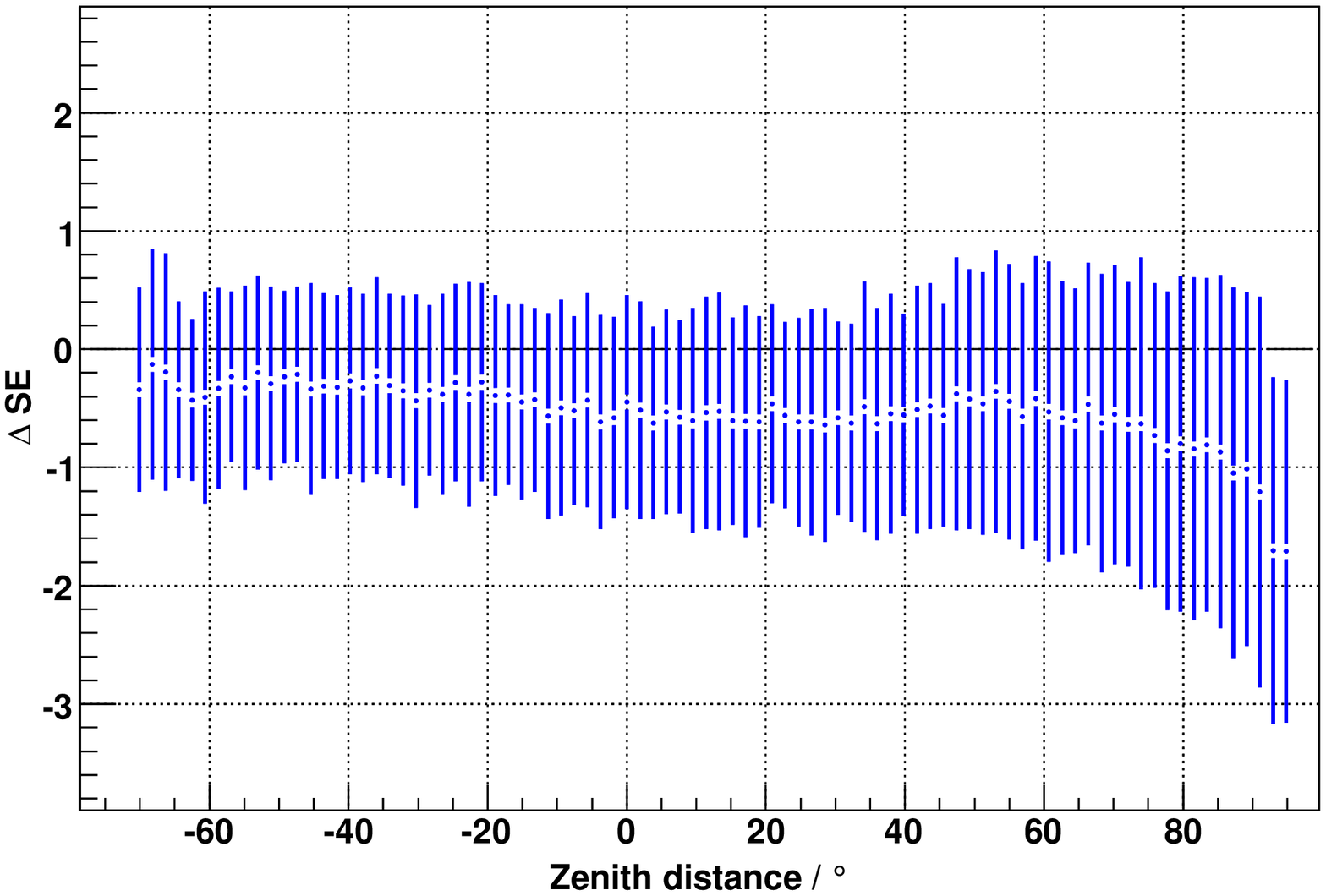}
\caption{The measured difference between the two shaft-encoders fixed
on either side of the elevation axis versus zenith angle. Negative
zenith angles mean that the telescope has been flipped over the zenith
position to the opposite side. The average offset from zero
corresponds to a the twist of the two shaft encoders with respect to
each other. The error bars denote the spread of several measurements. 
Under normal conditions the torsion between both ends of
the axis is less than the shaft-encoder resolution.}
\label{figure3}
\end{center}
\end{figure}

The angular telescope positions are measured by three shaft-encoders
(type designation Hengstler\footnote{\url{http://www.hengstler.de}\\Hengstler GmbH,
78554 Aldingen, Germany} AC61/1214EQ.72OLZ). These absolute multi-turn
encoders have a resolution of 4\,096 (10\,bit) revolutions and 16\,384
(14\,bit) steps per revolution, corresponding to an intrinsic angular
resolution of 1.3$^\prime$ per step. One shaft encoder is located on
the azimuth axis, while two more encoders are fixed on either side of
the elevation axis, increasing the resolution and allowing for
measurements of the twisting of the dish (fig.~\ref{figure3}). All
shaft encoders used are watertight (IP\,67) to withstand the extreme
weather conditions occasionally encountered at the telescope site. The
motor positions are read out at a frequency of 1\,kHz from 10\,bit
relative rotary encoders fixed on the motor axes. Due to the gear ratio
of more than one thousand between motor and load, the 14\,bit
resolution of the shaft encoder system on the axes can be interpolated
further using the position readout of the motors. For communication
with the axis encoders, a CANbus interface with the CANopen protocol is
in use (operated at 125\,kbps). The motor encoders are directly
connected by an analog interface.

\subsection{Motor control}

\begin{figure*}[htb]
\begin{center}
 \includegraphics*[width=0.78\textwidth,angle=0,clip]{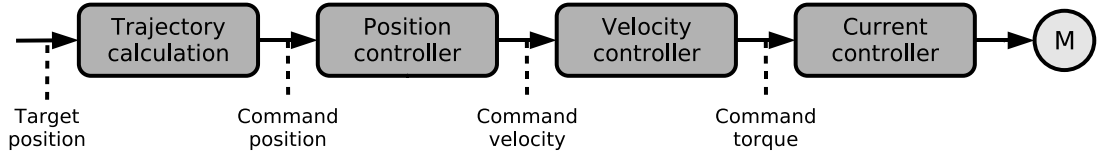}
 \caption{The internal flow control between the individual controllers
inside the drive control unit. Depending on the type of the command
value, different controllers are active. The control loops are closed by
the feedback of the rotary encoder on the motor, and a possible
controller on the load axis, as well as the measurement of the current.}
\label{figure4}
\end{center}
\end{figure*}

The three servo motors are connected to individual motion controller
units ({\em DKC}, type designation Bosch Rexroth,
DKC~ECODRIVE\,03.3-200-7-FW), serving as intelligent frequency
converters. An input value, given either analog or
digital, is converted to a predefined output, e.g., command position,
velocity or torque. All command values are processed through a chain of
built-in controllers, cf. fig.~\ref{figure4}, resulting in a final
command current applied to the motor. This internal chain of control
loops, maintaining the movement of the motors at a frequency of
1\,kHz, fed back by the rotary encoders on the corresponding motor axes.
Several safety limits ensure damage-free operation of the system even
under unexpected operation conditions. These safety limits are, e.g.,
software end switches, torque limits, current limits or
control-deviation limits.

To synchronize the two azimuth motors, a master-slave setup is
used. While the master is addressed by a command velocity, the
slave is driven by the command torque output of the master. This
operation mode ensures that both motors can apply their combined force
to the telescope structure without oscillations. In principle it is
possible to use a bias torque to eliminate play. This feature 
was not used because the play is negligible anyhow.

\subsection{Motion control}


The master for each axis is controlled by presetting a rotational
speed defined by $\pm$10\,V on its analog input. The input voltage is
produced by a programmable micro controller dedicated to analog motion
control, produced by Z\&B\footnote{\url{http://www.zub.de}\\zub machine control AG, 6074
Kastanienbaum, Switzerland} ({\em MACS}, type
designation MACS). The feedback is realized through a 500-step
emulation of the motor's rotary encoders by the DKCs. Elevation and azimuth movement is regulated by individual
MACSs. The MACS controller itself communicates with the control
software (see below) through a CANbus connection.

It turned out that in particular the azimuth motor system seems to be
limited by the large moment of inertia of the telescope
($J_{\mathrm{az}}$\,$\approx$\,4400\,tm$^2$, for comparison
$J_{\mathrm{el}}$\,$\approx$\,850\,tm$^2$; note that the exact numbers depend
on the current orientation of the telescope). At the same time, the
requirements on the elevation drive are much less demanding.\\

\noindent {\em MAGIC\,II}\quad For the drive system
several improvements have been provided:\\\vspace{-2ex}
\begin{itemize}
\item 13\,bit absolute shaft-encoders (type designation Heidenhain\footnote{\url{http://www.heidenhain.de}\\Dr.~Johannes Heidenhain GmbH, 83301 Traunreut, Germany}
ROQ\,425) are installed, providing an additional sine-shaped
$\pm$1\,Vss output within each step. This allows for a more accurate
interpolation and hence a better resolution than a simple 14\,bit
shaft-encoder. These shaft-encoders are also water tight (IP\,64), and
they are read out via an EnDat\,2.2 interface.
\item All encoders are directly connected to the DKCs, providing
additional feedback from the telescope axes itself. The DKC can control
the load axis additionally to the motor axis providing a more accurate
positioning, faster movement by improved oscillation suppression and a
better motion control of the system.
\item The analog transmission of the master's command torque to the
slave is replaced by a direct digital communication (EcoX)
of the DKCs. This allows for more robust and precise slave control.
Furthermore the motors could be coupled with relative angular synchronism
allowing to suppress deformations of the structure by keeping the
axis connecting both motors stable.
\item A single professional programmable logic controller (PLC), in German:
{\em Speicherprogammierbare Steuerung} (SPS, type designation Rexroth
Bosch, IndraControl SPS L\,20) replaces the two MACSs. Connection between
the SPS and the DKCs is now realized through a digital Profibus DP
interface substituting the analog signals.
\item The connection from the SPS to the control PC is realized via
Ethernet connection. Since Ethernet is more commonly in use than
CANbus, soft- and hardware support is much easier.
\end{itemize}

\subsection{PC control}

The drive system is controlled by a standard PC running a Linux
operating system, a custom-designed software based on
ROOT~\citep{www:root} and the positional astronomy library
{\em slalib}~\citep{slalib}.

Algorithms specialized for the MAGIC tracking system are imported from
the Modular Analysis and Reconstruction Software package
(MARS)~\citep{Bretz:2003icrc, Bretz:2005paris, Bretz:2008gamma} also
used in the data analysis~\citep{Bretz:2005mars, Dorner:2005paris}.

\subsubsection{Positioning}

Whenever the telescope has to be positioned, the relative distance to
the new position is calculated in telescope coordinates and then
converted to motor revolutions. Then, the micro controllers are
instructed to move the motors accordingly. Since the motion is
controlled by the feedback of the encoders on the motor axes, not on
the telescope axes, backlash and other non-deterministic irregularities
cannot easily be taken into account. Thus it may happen that the final
position is still off by a few shaft-encoder steps, although the motor
itself has reached its desired position. In this case, the procedure is
repeated up to ten times. After ten unsuccessful iterations, the system
would go into error state. In almost all cases the command position is
reached after at most two or three iterations.

If a slewing operation is followed by a tracking operation of a
celestial target position, tracking is started immediately after the
first movement without further iterations. Possible small deviations,
normally eliminated by the iteration procedure, are then corrected by
the tracking algorithm.

\subsubsection{Tracking}
To track a given celestial target position (RA/Dec, J\,2000.0,
FK\,5~\citep{1988VeARI..32....1F}), astrometric and misalignment
corrections have to be taken into account. While astrometric
corrections transform the celestial position into local coordinates as
seen by an ideal telescope (Alt/Az), misalignment corrections convert
them further into the coordinate system specific to the real telescope.
In case of MAGIC, this coordinate system is defined by the position
feedback system.

The tracking algorithm controls the telescope by applying a command
velocity for the revolution of the motors, which is re-calculated every
second. It is calculated from the current feedback position and the
command position required to point at the target five seconds ahead in
time. The timescale of 5\,s is a compromise between optimum tracking
accuracy and the risk of oscillations in case of a too short timescale.

As a crosscheck, the ideal velocities for the two telescope axes are
independently estimated using dedicated astrometric routines of slalib.
For security reasons, the allowable deviation between the determined
command velocities and the estimated velocities is limited. If an
extreme deviation is encountered the command velocity is set to zero,
i.e., the movement of the axis is stopped.

\subsection{Fast positioning}

The observation of GRBs and their afterglows in very-high energy
gamma-rays is a key science goal for the MAGIC telescope. Given that
alerts from satellite monitors provide GRB positions a few seconds
after their outburst via the {\em Gamma-ray Burst Coordination
Network}~\cite{www:gcn}, typical burst durations of 10\,s to
100\,s~\cite{Paciesas:1999} demand a fast positioning of the
telescope. The current best value for the acceleration has been set to
11.7\,mrad\,s$^{-2}$. It is constrained by the maximum constant force
which can be applied by the motors. Consequently, the maximum allowed
velocity can be derived from the distance between the end-switch
activation and the position at which a possible damage to the telescope
structure, e.g.\ ruptured cables, would happen. From these constraints,
the maximum velocity currently in use, 70.4\,mrad\,s$^{-1}$, was
determined. Note that, as the emergency stopping distance evolves
quadratically with the travel velocity, a possible increase of the
maximum velocity would drastically increase the required braking
distance. As safety procedures require, an emergency stop is completely
controlled by the DKCs itself with the feedback of the motor encoder,
ignoring all other control elements.

Currently, automatic positioning by
$\Delta\varphi$\,=\,180\textdegree{} in azimuth to the target position
is achieved within 45\,s. The positioning time in elevation is not
critical in the sense that the probability to move a longer path in
elevation than in azimuth is negligible. Allowing the telescope drive
to make use of the reverse mode, the requirement of reaching any
position in the sky within 30\,s is well met, as distances in azimuth
are substantially shortened. The motor specifications allow for a
velocity more than four times higher. In practice, the maximum possible
velocity is limited by the acceleration force, at slightly more than
twice the current value. The actual limiting factor is the braking
distance that allows a safe deceleration without risking any damage to
the telescope structure.

With the upgraded MAGIC\,II drive system, during commissioning in 2008 August, a
maximum acceleration and deceleration of $a_{az}$\,=\,30\,mrad\,s$^{-2}$
and $a_{zd}$\,=\,90\,mrad/s$^{-2}$ and a maximum velocity of
$v_{az}$\,=\,290\,mrad\,s$^{-1}$ and $v_{zd}$\,=\,330\,mrad\,s$^{-1}$
could be reached. With these values the limits of the motor power are
exhausted. This allowed a movement of
$\Delta\varphi$\,=\,180\textdegree/360\textdegree{} in azimuth within 20\,s\,/\,33\,s.

\subsection{Tracking precision}

The intrinsic mechanical accuracy of the tracking system is determined
by comparing the current command position of the axes with the feedback
values from the corresponding shaft encoders. These feedback values
represent the actual position of the axes with highest precision
whenever they change their feedback values. At these instances, the
control deviation is determined, representing the precision with which
the telescope axes can be operated. In the case of an ideal mount this
would define the tracking accuracy of the telescope.

\begin{figure}[htb]
\begin{center}
 \includegraphics*[width=0.48\textwidth,angle=0,clip]{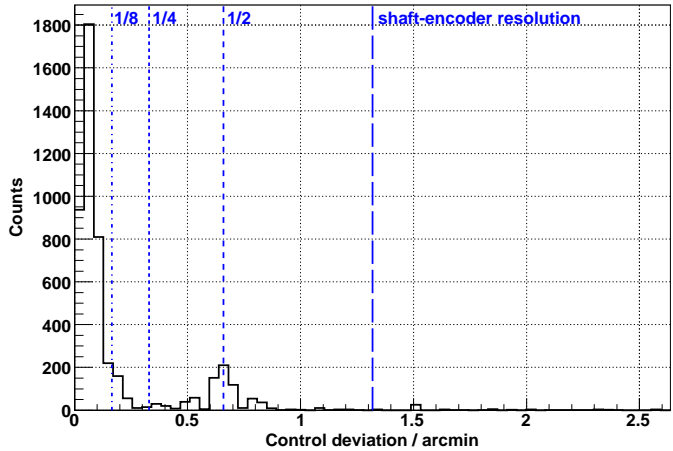}
 \caption{Control deviation between the expected, i.e. calculated,
position, and the feedback position of the shaft encoders in the moment
at which one change their readout values. For simplicity, the control
deviation is shown as absolute control deviation projected on the sky.
The blue lines correspond to fractions of the shaft-encoder resolution.
The peak at half of the shaft-encoder resolution results from cases
in which one of the two averaged elevation encoders is off by one step.
}
\label{figure5}
\end{center}
\end{figure}

In figure~\ref{figure5} the control deviation measured for 10.9\,h of
data taking in the night of 2007 July 22/23 and on the evening of July
23 is shown, expressed as absolute deviation on the sky taking both
axes into account. In almost all cases it is well below the resolution
of the shaft encoders, and in 80\% of the time it does not exceed 1/8
of this value ($\sim$10$^{\prime\prime}$). This means that the accuracy of the
motion control, based on the encoder feedback, is much better than
1$^\prime$ on the sky, which is roughly a fifth of the diameter of a
pixel in the MAGIC camera (6$^\prime$, c.f.~\cite{Beixeras:2005}).

In the case of a real telescope ultimate limits of the tracking
precision are given by the precision with which the correct command
value is known. Its calibration is discussed hereafter.

\section{Calibration}\label{sec4}

To calibrate the position command value, astrometric corrections
(converting the celestial target position into the target position of
an ideal telescope) and misalignment corrections (converting it further
into the target position of a real telescope), have to be taken into
account.

\subsection{Astrometric corrections}

\begin{figure}[htb]
\begin{center} 
 \includegraphics*[width=0.185\textwidth,angle=0,clip]{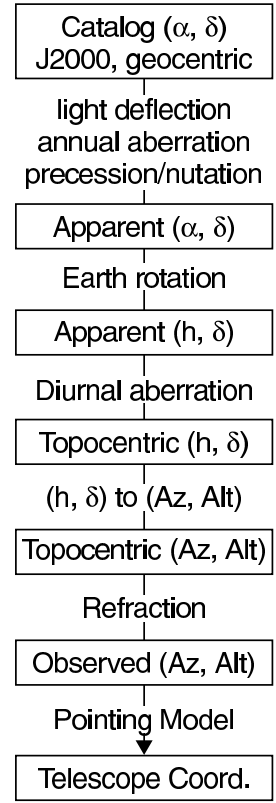}
\caption{The transformation applied to a given set of catalog source
coordinates to real-telescope coordinates. These corrections include
all necessary astrometric corrections, as well as the pointing
correction to transform from an ideal-telescope frame to the frame of a
real telescope. 
}
\label{figure6} 
\end{center} 
\end{figure}

The astrometric correction for the pointing and tracking algorithms is
based on a library for calculations usually needed in positional
astronomy, {\em slalib}~\cite{slalib}. Key features of this library are
the numerical stability of the algorithms and their well-tested
implementation. The astrometric corrections in use
(fig.~\ref{figure6}) -- performed when converting a celestial position
into the position as seen from Earth's center (apparent position) --
take into account precession and nutation of the Earth and annual
aberration, i.e., apparent displacements caused by the finite speed of
light combined with the motion of the observer around the Sun during
the year. Next, the apparent position is transformed to the observer's
position, taking into account atmospheric refraction, the Earth's
rotation, and diurnal aberration, i.e., the motion of the observer
around the Earth's rotation axis. Some of these effects are so small
that they are only relevant for nearby stars and optical astronomy. But
as optical observations of such stars are used to {\em train} the
misalignment correction, all these effects are taken into account.

\subsection{Pointing model}

Imperfections and deformations of the mechanical construction lead to
deviations from an ideal telescope, including the non-exact
alignment of axes, and deformations of the telescope structure. 

In the case of the MAGIC telescopes the optical axis of the mirror is
defined by an automatic alignment system. This active mirror control
is programmed not to change the optical axis once defined, but only
controls the optical point spread function of the mirror, i.e., it does
not change the center of gravity of the light distribution. 
This procedure is applied whenever the telescope is observing including
any kind of calibration measurement for the drive system. The precision
of the axis alignment of the mirrors is better than 0.2$^\prime$ and can
therefor be neglected.

Consequently, to assure reliable pointing and tracking accuracy, 
mainly the mechanical effects have to be taken into account. 
Therefore the tracking software employs an analytical pointing model
based on the {\rm TPOINT}\texttrademark{} telescope modeling
software~\cite{tpoint}, also used for optical telescopes. This model,
called {\em pointing model}, parameterizes deviations from the ideal
telescope. Calibrating the pointing model by mispointing measurements
of bright stars, which allows to determine the necessary corrections,
is a standard procedure. Once calibrated, the model is applied online.
Since an analytical model is used, the source of any deviation can be
identified and traced back to components of the telescope mount.\\

Corrections are parameterized by alt-azimuthal terms~\cite{tpoint},
i.e., derived from vector transformations within the proper coordinate
system. The following possible misalignments are taken into account:\\\vspace{-2ex}
\begin{description}
\item[Zero point corrections ({\em index errors})] Trivial offsets
between the zero positions of the axes and the zero positions of the
shaft encoders.
\item[Azimuth axis misalignment] The misalignment of the azimuth axis
in north-south and east-west direction, respectively. For MAGIC these
corrections can be neglected.
\item[Non-perpendicularity of axes] Deviations from right angles
between any two axes in the system, namely (1) non-perpendicularity of
azimuth and elevation axes and (2) non-perpendicularity of elevation
and pointing axes. In the case of the MAGIC telescope these corrections
are strongly bound to the mirror alignment defined by the active mirror
control.
\item[Non-centricity of axes] The once-per-revolution cyclic errors
produced by de-centered axes. This correction is small, and thus difficult
to measure, but the most stable correction throughout the years.
\end{description}

\noindent{\bf Bending of the telescope structure}
\begin{itemize}
\item A possible constant offset of the mast bending.
\item A zenith angle dependent correction. It describes the camera mast
bending, which originates by MAGIC's single thin-mast camera support
strengthened by steel cables.
\item Elevation hysteresis: This is an offset correction introduced
depending on the direction of movement of the elevation axis. It is
necessary because the sliding bearing, having a stiff connection with
the encoders, has such a high static friction that in case of reversing
the direction of the movement, the shaft-encoder will not indicate any
movement for a small and stable rotation angle, even though the
telescope is rotating. Since this offset is stable, it can easily 
be corrected after it is fully passed. The passage of the hysteresis
is currently corrected offline only.
\end{itemize}
\vspace{1em}

Since the primary feedback is located on the axis itself, corrections
for irregularities of the chain mounting or sprocket wheels are
unnecessary. Another class of deformations of the telescope-support
frame and the mirrors are non-deterministic and, consequently, pose an
ultimate limit of the precision of the pointing.


\subsection{Determination}


To determine the coefficients of a pointing model, calibration
data is recorded. It consists of mispointing measurements depending
on altitude and azimuth angle. Bright stars are tracked with the
telescope at positions uniformly distributed in local coordinates,
i.e., in altitude and azimuth angle. The real pointing
position is derived from the position of the reflection of a bright
star on a screen in front of the MAGIC camera. The center
of the camera is defined by LEDs mounted on an ideal ($\pm$1\,mm)
circle around the camera center, cf.~\citet{Riegel:2005icrc}.

Having enough of these datasets available, correlating ideal and real
pointing position, a fit of the coefficients of the model can be made,
minimizing the pointing residual.

\subsubsection{Hardware and installations}

\begin{figure*}[htb]
\begin{center}
 \includegraphics*[width=0.48\textwidth,angle=0,clip]{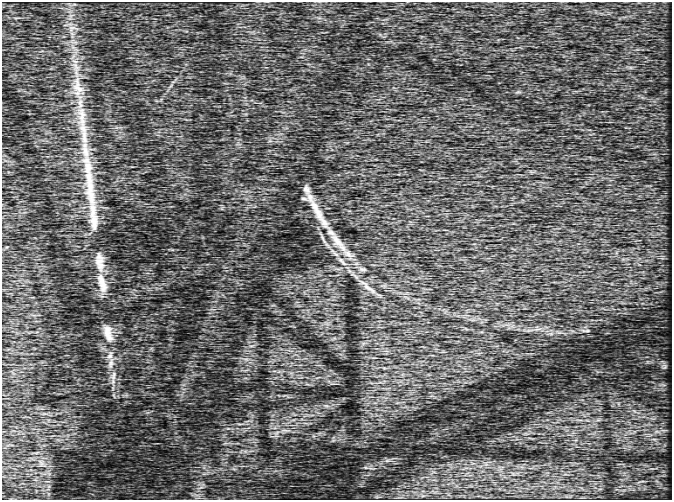}
 \hfill
 \includegraphics*[width=0.48\textwidth,angle=0,clip]{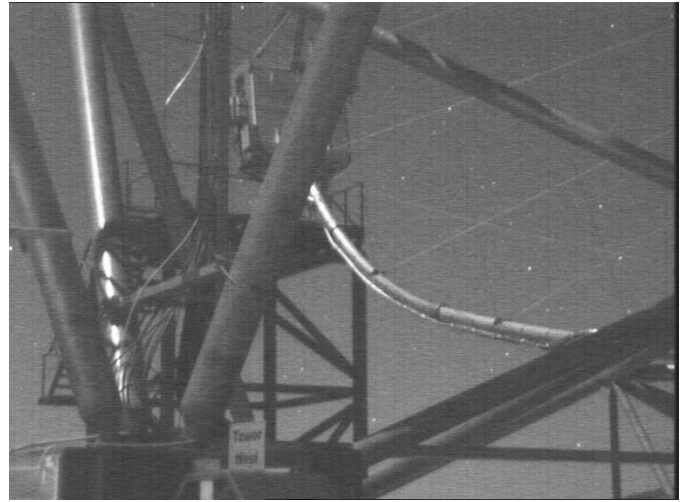}
 \caption{A single frame (left) and an average of 125 frames (right) of
the same field of view taken with the high sensitivity PAL CCD camera
used for calibration of the pointing model. The frames were taken
with half moon. }
\label{figure7}
\end{center}
\end{figure*}

A 0.0003\,lux, 1/2\(^{\prime\prime}\) high-sensitivity standard PAL CCD
camera (type designation \mbox{Watec}~WAT-902\,H) equipped with a zoom lens (type: Computar) is used
for the mispointing measurements. The camera is read out at a rate of
25\,frames per second using a standard frame-grabber card in a standard
PC. The camera has been chosen providing adequate performance and
easy readout, due to the use of standard components, for a very cheap price
($<$\,500\,Euro). The tradeoff for the high sensitivity of the camera is its high
noise level in each single frame recorded. Since there are no rapidly
moving objects within the field of view, a high picture quality can be
achieved by averaging typically 125\,frames (corresponding to 5\,s). An
example is shown in figure~\ref{figure7}. This example also illustrates
the high sensitivity of the camera, since both pictures of the
telescope structure have been taken with the residual light of less
than a half-moon. In the background individual stars can be seen.
Depending on the installed optics, stars up to 12$^\mathrm{m}$ are
visible. With our optics and a safe detection threshold the limiting
magnitude is typically slightly above 9$^\mathrm{m}$ for direct
measurements and on the order of 5$^\mathrm{m}$\dots4$^\mathrm{m}$ for
images of stars on the screen.

\subsubsection{Algorithms}

\begin{figure}[htb]
\begin{center}
 \includegraphics*[width=0.48\textwidth,angle=0,clip]{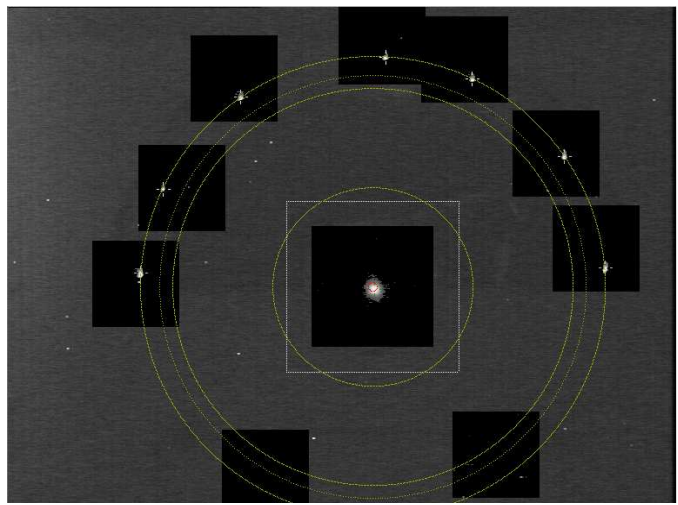}
\caption{A measurement of a star for the calibration of the pointing
model. Visible are the seven LEDs and their determined center of
gravity, as well as the reconstructed circle on which the LEDs are
located. The LEDs on the bottom part are hidden by the lower lid,
holding a screen in front of the camera. 
The black regions are the search regions for
the LEDs and the calibration star. A few dead pixels in the CCD camera
can also be recognized.}
\label{figure8}
\end{center}
\end{figure}

An example of a calibration-star measurement is shown in
figure~\ref{figure8}. Using the seven LEDs mounted on a circle around
the camera center, the position of the camera center is determined.
Only the upper half of the area instrumented is visible, since
the lower half is covered by the lower lid, holding a special
reflecting surface in the center of the camera. The LED positions are
evaluated by a simple cluster-finding algorithm looking at pixels more
than three standard deviations above the noise level. The LED position
is defined as the center of gravity of its light distribution, its
search region by the surrounding black-colored boxes. For
simplicity the noise level is determined just by calculating the mean
and the root-mean-square within the individual search regions below a
fixed threshold dominated by noise.

Since three points are enough to define a circle, from all
possible combinations of detected spots, the corresponding circle
is calculated. In case of misidentified LEDs, which sometimes
occur due to light reflections from the telescope structure, the
radius of the circle will deviate from the predefined radius.
Thus, any such misidentified circles are discarded. The radius
determination can be improved further by applying small offsets of
the non-ideal LED positions. The radius distribution is Gaussian
and its resolution is $\sigma$\,$\apprle$\,1\,mm
($\mathrm{d}r/r\approx0.3$\textperthousand) on the camera plane
corresponding to $\sim$1$^{\prime\prime}$.

The center of the ring is calculated as the average of all circle
centers after quality cuts. Its resolution is
$\sim$2$^{\prime\prime}$. In this setup, the large number of LEDs guarantees
operation even in case one LED could not be detected due to damage or
scattered light.

To find the spot of the reflected star itself, the same cluster-finder
is used to determine its center of gravity. This gives reliable results
even in case of saturation. Only very bright stars, brighter than
1.0$^m$, are found to saturate the CCD camera asymmetrically.

Using the position of the star, with respect to the camera center, the
pointing position corresponding to the camera center is calculated.
This position is stored together with the readout from the position
feedback system. The difference between the telescope pointing
position and the feedback position is described by the pointing model.
Investigating the dependence of these differences on zenith and azimuth
angle, the correction terms of the pointing model can be determined.
Its coefficients are fit minimizing the absolute residuals on the celestial
sphere.

\subsection{Results}

\begin{figure}[htb]
\begin{center}
 \includegraphics*[width=0.48\textwidth,angle=0,clip]{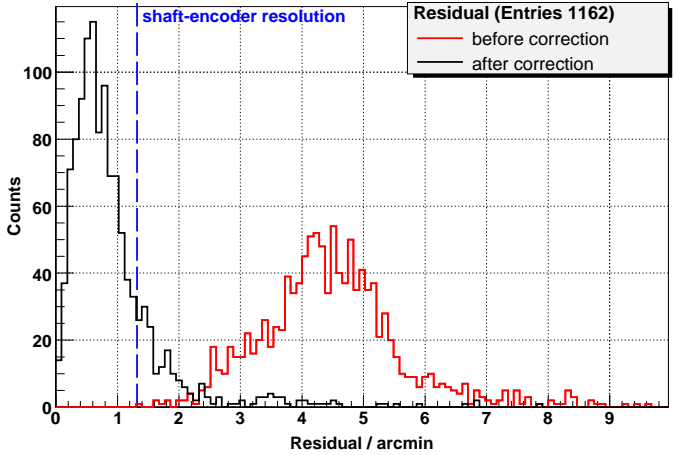}
 \caption{Distribution of absolute pointing residual on the sky between
the measured position of calibration stars and their nominal position
with only offset correction for both axes (red) and a fitted pointing
model (blue) applied. Here in total 1162 measurements where used,
homogeneously distributed over the local sky. After application of the
pointing model the residuals are well below the shaft-encoder
resolution, i.e., the knowledge of the mechanical position of the axes.
\label{figure9}
}

\end{center}
\end{figure}

Figure~\ref{figure9} shows the residuals, taken between 2006 October and
2007 July, before and after application of the fit of the pointing
model. For convenience, offset corrections are applied to the residuals
before correction. Thus, the red curve is a measurement of the alignment
quality of the structure, i.e., the pointing accuracy with offset
corrections only. By fitting a proper model, the pointing accuracy can
be improved to a value below the intrinsic resolution of the system,
i.e., below shaft-encoder resolution. In more than 83\% of all cases the
tracking accuracy is better than 1.3$^\prime$
and it hardly ever exceeds 2.5$^\prime$. The few datasets exceeding
2.5$^\prime$ are very likely due to imperfect measurement of the
real pointing position of the telescope, i.e., the center of gravity of
the star light.

The average absolute correction applied (excluding the index error) is on
the order of 4$^\prime$. Given the size, weight and structure of
the telescope this proves a very good alignment and low sagging of the
structure. The elevation hysteresis, which is intrinsic to the
structure, the non-perpendicularity and non-centricity of the axes are
all in the order of 3$^\prime$, while the azimuth axis
misalignment is $<$\,0.6$^\prime$. These numbers are well in agreement
with the design tolerances of the telescope.

\subsubsection{Limitations}

The ultimate limit on the achievable pointing precision are effects,
which are difficult to correlate or measure, and non-deterministic
deformations of the structure or mirrors. For example, the azimuth
support consists of a railway rail with some small deformations in
height due to the load, resulting in a wavy movement difficult to
parameterize. For the wheels on the six bogeys, simple, not precisely
machined crane wheels were used, which may amplify horizontal
deformations. Other deformations are caused by temperature changes and
wind loads which are difficult to control for telescopes without dome,
and which cannot be modeled. It should be noted that the azimuth structure
can change its diameter by up to 3\,cm due to day-night temperature
differences, indicating that thermal effects have a non-negligible and
non-deterministic influence.

Like every two axis mount, also an alt-azimuth mount has a blind spot
near its upward position resulting from misalignments of the axis which
are impossible to correct by moving one axis or the other. From the
size of the applied correction it can be derived that the blind spot
must be on the order of $\lesssim$\,6$^\prime$ around zenith. 
Although the MAGIC drive system is powerful enough to keep on track
pointing about 6$^\prime$ away from zenith, for safety reasons, i.e.,
to avoid fast movment under normal observation conditions, the observation
limit has been set to $\theta$\,$<$\,30$^\prime$. Such fast movements
are necessary to change the azimuth position from moving the telescope 
upwards in the East to downwards in the South. In the case of an ideal
telescope, pointing at zenith, even an infinitely fast movement would
be required.

\subsubsection{Stability}

\begin{figure}[htb]
\begin{center}
 \includegraphics*[width=0.48\textwidth,angle=0,clip]{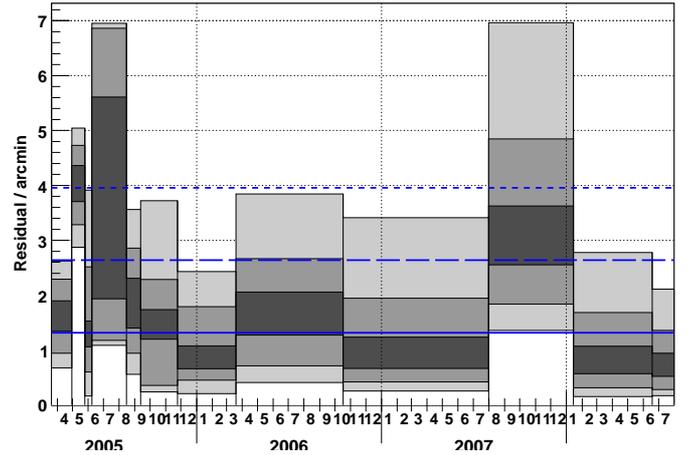}
\caption{The distribution of mispointing measurements. The
measurement is a direct measurement of the pointing accuracy. The plot
shows its time-evolution. Details on the bin edges and the available
statistics is given in the caption of table~\ref{table2}. Since the
distribution is asymmetric, quantiles are shown, from bottom to top, at
5\%, 13\%, 32\%, 68\%, 87\% and 95\%. The dark grey region belong to
the region between quantiles 32\% and 68\%. } 
\label{figure10}
\end{center} 
\end{figure}

With each measurement of a calibration-star also the present pointing
uncertainty is recorded. This allows for monitoring of the pointing
quality and for offline correction. In figure~\ref{figure10} the
evolution of the measured residuals over the years are shown. The
continuous monitoring has been started in March 2005 and is still
ongoing. Quantiles are shown since the distribution can be
asymmetric depending on how the residuals are distributed on the sky. The
points have been grouped, where the grouping reflects data taken under
the same conditions (pointing model, mirror alignment, etc.). It should
be noted, that the measured residuals depend on zenith and azimuth
angle, i.e., the distributions shown are biased due to inhomogeneous
distributions on the sky in case of low statistics. Therefore the
available statistics is given in table~\ref{table2}. 
\begin{table}[htb]
\begin{center}
\begin{tabular}{|l|c|}\hline
Begin&Counts\\\hline\hline
2005/03/20&29\\
2005/04/29&43\\
2005/05/25&30\\
2005/06/08&26\\
2005/08/15&160\\
2005/09/12&22\\\hline
\end{tabular}
\hfill
\begin{tabular}{|l|c|}\hline
Begin&Counts\\\hline\hline
2005/11/24&38\\
2006/03/19&502\\
2006/10/17&827\\
2007/07/31&87\\
2008/01/14&542\\
2008/06/18&128\\\hline
\end{tabular}\hfill
\end{center}
\caption{Available statistics corresponding to the distributions
shown in figure~\ref{figure10}. Especially in cases of low statistics
the shown distribution can be influenced by inhomogeneous distribution
of the measurement on the local sky. The dates given correspond to dates
for which a change in the pointing accuracy, as for example a change to
the optical axis or the application of a new pointing model, is known.}
\label{table2}
\end{table}

The mirror focusing can influence the alignment of the optical axis of
the telescope, i.e., it can modify the pointing model. Therefore a
calibration of the mirror refocusing can worsen the tracking accuracy,
later corrected by a new pointing model. Although the automatic mirror 
control is programmed such that a new calibration should not change the
center of gravity of the light distribution, it happened sometimes in
the past due to software errors. 

The determination of the pointing model also relies on a good
statistical basis, because the measured residuals are of a similar
magnitude as the accuracy of a single calibration-star measurement. The
visible improvements and deterioration are mainly a consequence of new
mirror focusing and following implementations of new pointing models.
The improvement over the past year is explained by the gain in
statistics.

On average the systematic pointing uncertainty was always better than
three shaft-encoder steps (corresponding to 4$^\prime$), most of the
time better than 2.6$^\prime$ and well below one shaft-encoder step,
i.e.\ 1.3$^\prime$, in the past year. Except changes to the pointing
model or the optical axis, as indicated by the bin edges, no
degradation or change with time of the pointing model or
a worsening of the limit given by the telescope mechanics could be found.

\section{Scalability}\label{sec5}

With the aim to reach lower energy thresholds, the next generation of
Cherenkov telescopes will also include larger and heavier ones. 
Therefore more powerful drive systems will be needed. The scalable
drive system of the MAGIC telescope is suited to meet this challenge.
With its synchronous motors and their master-slave setup, it can easily
be extended to larger telescopes at moderate costs, or even scaled down
to smaller ones using less powerful components. Consequently,
telescopes in future projects, with presumably different sizes, can be
driven by similar components resulting in a major 
simplification of maintenance. With the current setup, a tracking
accuracy at least of the order of the shaft-encoder resolution is
guaranteed. Pointing accuracy -- already including all possible
pointing corrections -- is dominated by dynamic and unpredictable
deformations of the mount, e.g., temperature expansion.

\section{Outlook}\label{outlook}

Currently, efforts are ongoing to implement the astrometric subroutines
as well as the application of the pointing model directly into the
Programmable Logic Controller. A first test will be carried out within
the DWARF project soon~\cite{DWARF}. The direct advantage is that the
need for a control PC is omitted. Additionally, with a more direct
communication between the algorithms, calculating the nominal position
of the telescope mechanics, and the control loop of the drive
controller, a real time, and thus more precise, position control can be
achieved. As a consequence, the position controller can directly be
addressed, even when tracking, and the outermost position control-loop
is closed internally in the drive controller. This will ensure an even
more accurate and stable motion. Interferences from external sources,
e.g. wind gusts, could be counteracted at the moment of appearance by
the control on very short timescales, on the order of milli-seconds. An
indirect advantage is that with a proper setup of the control loop
parameters, such a control is precise and flexible enough that a
cross-communication between the master and the slaves can also be
omitted. Since all motors act as their own master, in such a system a
broken motor can simply be switched off or mechanically decoupled
without influencing the general functionality of the system.

An upgrade of the MAGIC\,I drive system according to the improvements 
applied for MAGIC\,II is ongoing.

\section{Conclusions}\label{conclusions}

The scientific requirements demand a powerful, yet accurate drive
system for the MAGIC telescope. From its hardware installation and
software implementation, the installed drive system exceeds its design
specifications as given in section~\ref{design}. At the same time the
system performs reliably and stably, showing no deterioration after
five years of routine operation. The mechanical precision of the motor
movement is almost ten times better than the readout on the telescope
axes. The tracking accuracy is dominated by random deformations and
hysteresis effects of the mount, but still negligible 
compared to the measurement of the position of the telescope axes. The
system features integrated tools, like an analytical pointing model.
Fast positioning for gamma-ray burst followup is achieved on average
within less than 45 seconds, or, if movements {\em across the zenith}
are allowed, 30 seconds. Thus, the drive system makes
MAGIC the best suited telescope for observations of these phenomena at
very high energies.

For the second phase of the MAGIC project and particularly for the
second telescope, the drive system has been further improved.
By design, the drive system is easily scalable from its current
dimensions to larger and heavier telescope installations as required
for future projects. The improved stability is also expected to meet
the stability requirements, necessary when operating a larger number of
telescopes.

\section[]{Acknowledgments}
The authors acknowledge the support of the MAGIC collaboration, and
thank the IAC for providing excellent working conditions at the
Observatorio del Roque de los Muchachos. The MAGIC project is mainly
supported by BMBF (Germany), MCI (Spain), INFN (Italy). We thank the
construction department of the MPI for Physics, for their help in the
design and installation of the drive system as well as Eckart Lorenz,
for some important comments concerning this manuscript. R.M.W.\
acknowledges financial support by the MPG. His research is also
supported by the DFG Cluster of Excellence ``Origin and Structure of
the Universe''.

\end{document}